\documentclass[conference]{IEEEtran}[conference]

\newcommand{\pdftitle}{Radio Environment Mapping with World Models for Active Measurement Control: Should Networks Dream of Optimal Control?}

\newcommand\copyrighttext{%
  \footnotesize \textcopyright 2026 IEEE. Personal use of this material is permitted.
  Permission from IEEE must be obtained for all other uses, in any current or future
  media, including reprinting/republishing this material for advertising or promotional
  purposes, creating new collective works, for resale or redistribution to servers or
  lists, or reuse of any copyrighted component of this work in other works. This paper was accepted for presentation at at EuCNC (European Conference on Networks and Communications) \& 6G Summit 2026..
  }
\newcommand\copyrightnotice{%
\begin{tikzpicture}[remember picture,overlay]
\node[anchor=north,yshift=-5pt] at (current page.north) {\fbox{\parbox{\dimexpr\textwidth-\fboxsep-\fboxrule\relax}{\copyrighttext}}};
\end{tikzpicture}%
}

\usepackage[nolist,nohyperlinks]{acronym}
\usepackage{amsmath}
\usepackage{breqn}
\usepackage{balance}
\usepackage{cite}
\usepackage{framed}
\usepackage{siunitx}
\usepackage{soul}
\usepackage{multirow}
\usepackage{etoolbox}
\usepackage{standalone}
\usepackage{lipsum}

\usepackage{blindtext}

\usepackage{graphicx}
\usepackage{subcaption}
\usepackage{pgfplots}
\usepgfplotslibrary{groupplots}
\usepackage{wrapfig}
\usepackage{color}
\usepackage[table,xcdraw]{xcolor}

\graphicspath{{./images/}}

\usepackage{amsfonts} 

\usepackage{tikz}
\usetikzlibrary{arrows.meta, positioning, shapes.multipart, fit, calc}
\usepackage{algorithm}
\usepackage{algpseudocode}

\usepackage{url}
\usepackage[bookmarks=false]{hyperref}
\hypersetup{
  backref,
  hidelinks,
  colorlinks=false,
  breaklinks=true,
  bookmarksopen=false,
  pdftitle=\pdftitle,
  pdfauthor={Author}
}

\bstctlcite{IEEEexample:BSTcontrol}

\newtoggle{authorcomment}
\toggletrue{authorcomment}

\begin{document}
\bstctlcite{IEEEexample:BSTcontrol}

\title{\pdftitle}

\author{\IEEEauthorblockN{Jernej Hribar\IEEEauthorrefmark{1},
Ljupcho Milosheski\IEEEauthorrefmark{1}\IEEEauthorrefmark{2} and 
Ryoichi Shinkuma\IEEEauthorrefmark{3}}
\IEEEauthorblockA{\IEEEauthorrefmark{1}Department for Communication Systems, Jožef Stefan Institute, Ljubljana, Slovenia\\
\IEEEauthorrefmark{2}Jožef Stefan International Postgraduate School, Ljubljana, Slovenia\\
\IEEEauthorrefmark{3}Faculty of Engineering, Shibaura Institute of
Technology, Tokyo, Japan\\
Email: \{jernej.hribar, ljupcho.milosheski\}@ijs.si }}

\maketitle

\begin{acronym}[MACHU]
  \acro{iot}[IoT]{Internet of Things}
  \acro{cr}[CR]{Cognitive Radio}
  \acro{ofdm}[OFDM]{orthogonal frequency-division multiplexing}
  \acro{ofdma}[OFDMA]{orthogonal frequency-division multiple access}
  \acro{scfdma}[SC-FDMA]{single carrier frequency division multiple access}
  \acro{rbi}[RBI]{ Research Brazil Ireland}
  \acro{rfic}[RFIC]{radio frequency integrated circuit}
  \acro{sdr}[SDR]{Software Defined Radio}
  \acro{sdn}[SDN]{Software Defined Networking}
  \acro{su}[SU]{Secondary User}
  \acro{ra}[RA]{Resource Allocation}
  \acro{qos}[QoS]{quality of service}
  \acro{usrp}[USRP]{Universal Software Radio Peripheral}
  \acro{mno}[MNO]{Mobile Network Operator}
  \acro{mnos}[MNOs]{Mobile Network Operators}
  \acro{gsm}[GSM]{Global System for Mobile communications}
  \acro{tdma}[TDMA]{Time-Division Multiple Access}
  \acro{fdma}[FDMA]{Frequency-Division Multiple Access}
  \acro{gprs}[GPRS]{General Packet Radio Service}
  \acro{msc}[MSC]{Mobile Switching Centre}
  \acro{bsc}[BSC]{Base Station Controller}
  \acro{umts}[UMTS]{universal mobile telecommunications system}
  \acro{Wcdma}[WCDMA]{Wide-band code division multiple access}
  \acro{wcdma}[WCDMA]{wide-band code division multiple access}
  \acro{cdma}[CDMA]{code division multiple access}
  \acro{lte}[LTE]{Long Term Evolution}
  \acro{papr}[PAPR]{peak-to-average power rating}
  \acro{hn}[HetNet]{heterogeneous networks}
  \acro{phy}[PHY]{physical layer}
  \acro{mac}[MAC]{medium access control}
  \acro{amc}[AMC]{adaptive modulation and coding}
  \acro{mimo}[MIMO]{multiple input multiple output}
  \acro{rats}[RATs]{radio access technologies}
  \acro{vni}[VNI]{visual networking index}
  \acro{rbs}[RB]{resource blocks}
  \acro{rb}[RB]{resource block}
  \acro{ue}[UE]{user equipment}
  \acro{cqi}[CQI]{Channel Quality Indicator}
  \acro{hd}[HD]{half-duplex}
  \acro{fd}[FD]{full-duplex}
  \acro{sic}[SIC]{self-interference cancellation}
  \acro{si}[SI]{self-interference}
  \acro{bs}[BS]{base station}
  \acro{fbmc}[FBMC]{Filter Bank Multi-Carrier}
  \acro{ufmc}[UFMC]{Universal Filtered Multi-Carrier}
  \acro{scm}[SCM]{Single Carrier Modulation}
  \acro{isi}[ISI]{inter-symbol interference}
  \acro{ftn}[FTN]{Faster-Than-Nyquist}
  \acro{m2m}[M2M]{machine-to-machine}
  \acro{mtc}[MTC]{machine type communication}
  \acro{mmw}[mmWave]{millimeter wave}
  \acro{bf}[BF]{beamforming}
  \acro{los}[LOS]{line-of-sight}
  \acro{nlos}[NLOS]{non line-of-sight}
  \acro{capex}[CAPEX]{capital expenditure}
  \acro{opex}[OPEX]{operational expenditure}
  \acro{ict}[ICT]{information and communications technology}
  \acro{sp}[SP]{service providers}
  \acro{inp}[InP]{infrastructure providers}
  \acro{mvnp}[MVNP]{mobile virtual network provider}
  \acro{mvno}[MVNO]{mobile virtual network operator}
  \acro{nfv}[NFV]{network function virtualization}
  \acro{vnfs}[VNF]{virtual network functions}
  \acro{cran}[C-RAN]{Cloud Radio Access Network}
  \acro{bbu}[BBU]{baseband unit}
  \acro{bbus}[BBU]{baseband units}
  \acro{rrh}[RRH]{remote radio head}
  \acro{rrhs}[RRH]{Remote radio heads} 
  \acro{sfv}[SFV]{sensor function virtualization}
  \acro{wsn}[WSN]{wireless sensor networks} 
  \acro{bio}[BIO]{Bristol is open}
  \acro{vitro}[VITRO]{Virtualized dIstributed plaTfoRms of smart Objects}
  \acro{os}[OS]{operating system}
  \acro{www}[WWW]{world wide web}
  \acro{iotvn}[IoT-VN]{IoT virtual network}
  \acro{mems}[MEMS]{micro electro mechanical system}
  \acro{mec}[MEC]{Mobile edge computing}
  \acro{coap}[CoAP]{Constrained Application Protocol}
  \acro{vsn}[VSN]{Virtual sensor network}
  \acro{rest}[REST]{REpresentational State Transfer}
  \acro{aoi}[AoI]{Age of Information}
  \acro{lora}[LoRa\texttrademark]{Long Range}
  \acro{iot}[IoT]{Internet of Things}
  \acro{snr}[SNR]{Signal-to-Noise Ratio}
  \acro{cps}[CPS]{Cyber-Physical System}
  \acro{uav}[UAV]{Unmanned Aerial Vehicle}
  \acro{rfid}[RFID]{Radio-frequency identification}
  \acro{lpwan}[LPWAN]{Low-Power Wide-Area Network}
  \acro{lgfs}[LGFS]{Last Generated First Served}
  \acro{wsn}[WSN]{wireless sensor network} 
  \acro{lmmse}[LMMSE]{Linear Minimum Mean Square Error}
  \acro{rl}[RL]{Reinforcement Learning}
  \acro{nb-iot}[NB-IoT]{Narrowband IoT}
  \acro{lorawan}[LoRaWAN]{Long Range Wide Area Network}
  \acro{mdp}[MDP]{Markov Decision Process}
  \acro{ann}[ANN]{Artificial Neural Network}
  \acro{dqn}[DQN]{Deep Q-Network}
  \acro{mse}[MSE]{Mean Square Error}
  \acro{ml}[ML]{Machine Learning}
  \acro{cpu}[CPU]{Central Processing Unit}
  \acro{ddpg}[DDPG]{Deep Deterministic Policy Gradient}
  \acro{ai}[AI]{Artificial Intelligence}
  \acro{gp}[GP]{Gaussian Processes}
  \acro{drl}[DRL]{Deep Reinforcement Learning}
  \acro{mmse}[MMSE]{Minimum Mean Square Error}
  \acro{fnn}[FNN]{Feedforward Neural Network}
  \acro{eh}[EH]{Energy Harvesting}
  \acro{wpt}[WPT]{Wireless Power Transfer}
  \acro{dl}[DL]{Deep Learning}
  \acro{yolo}[YOLO]{You Only Look Once}
  \acro{mec}[MEC]{Mobile Edge Computing}
  \acro{marl}[MARL]{Multi-Agent Reinforcement Learning}
  \acro{rssi}[RSSI]{Received Signal Strength Indicator}
  \acro{ap}[AP]{Access Point}
  \acro{rem}[REM]{Radio Environment Mapping}
  \acro{rmse}[RMSE]{Root Mean Square Error}
  \acro{vae}[VAE]{Variational Autoencoder}
  \acro{lstm}[LSTM]{Long Short-term Memory}
  \acro{mae}[MAE]{Mean Absolute Error}
  \acro{gan}[GAN]{Generative Adversarial Network}
  \acro{rss}[RSS]{Radio Signal Straight}
\end{acronym}

\begin{abstract}
Radio Environment Maps (REMs) have the potential to serve as an important enabler for intelligent modeling and control in emerging AI-native 6G networks. Despite significant progress, most REM construction methods remain passive, relying on interpolation or static uncertainty models and lacking an explicit mechanism to reason about how future measurements will affect reconstruction quality under a limited measurement budget. In this paper, we formulate REM construction as a sequential decision-making problem and propose a world-model–inspired framework for active Received Signal Strength Indicator (RSSI) map reconstruction. By learning an internal representation of the radio environment and employing a dreaming mechanism to simulate the impact of candidate measurements, the proposed approach actively selects measurement locations under a limited budget. Experimental results on real indoor RSSI data demonstrate that the proposed method significantly outperforms Gaussian Process–based interpolation in the few-shot regime, achieving up to a fivefold reduction in Root Mean Square Error (RMSE) with the same number of measurements. These results highlight the potential of world models as a powerful paradigm for sample-efficient radio environment mapping and intelligent model-based sensing in 6G and beyond networks.

\end{abstract}

\acresetall

\begin{IEEEkeywords}
Radio Environment Maps, World Models, Machine Learning
\end{IEEEkeywords}

\copyrightnotice

\section{Introduction}
\label{sec:intro}





In emerging AI-native 6G networks, \acp{rem} are expected to play an important role in enabling environment-aware modeling and network optimization~\cite{zeng2021toward}. Applications such as indoor localization, network planning, spectrum management, and digital twins~\cite{bi2019engineering} can improve performance by employing accurate and easy-to-obtain \acp{rem}. Furthermore, in indoor environments, such radio maps containing information regarding wireless coverage can help reduce malfunctions or improve decision-making for many automation tasks, e.g., robots navigating in smart buildings. Unfortunately, obtaining these maps with a limited number of measurements, while also accounting for dynamic changes such as moving objects or users in the environment, remains a challenge.



In deployment scenarios, measurements are also often sparse or costly to obtain. To overcome this challenge, a wide range of classical approaches for constructing \acp{rem} have been proposed, including interpolation and model based techniques such as \ac{gp}, Kriging, etc.,~\cite{pesko2014radio}, as well as deep learning-based methods using convolution layers and \acp{gan}~\cite{feng2025recent}.  While these methods can yield accurate radio map estimates under static conditions or dense measurements, their performance degrades when the environment changes or when only a limited number of measurements is available. In particular, they lack an explicit mechanism to reason about how new measurements will affect future predictions and therefore cannot actively guide measurement acquisition. To that end, we employ world model framework~\cite{ha2018recurrent}, which learns an internal representation of the environment and enable model-based reasoning about the expected impact of future measurements.

\begin{figure*}[t!]
    \centering  
    \includegraphics[width=0.95\textwidth]{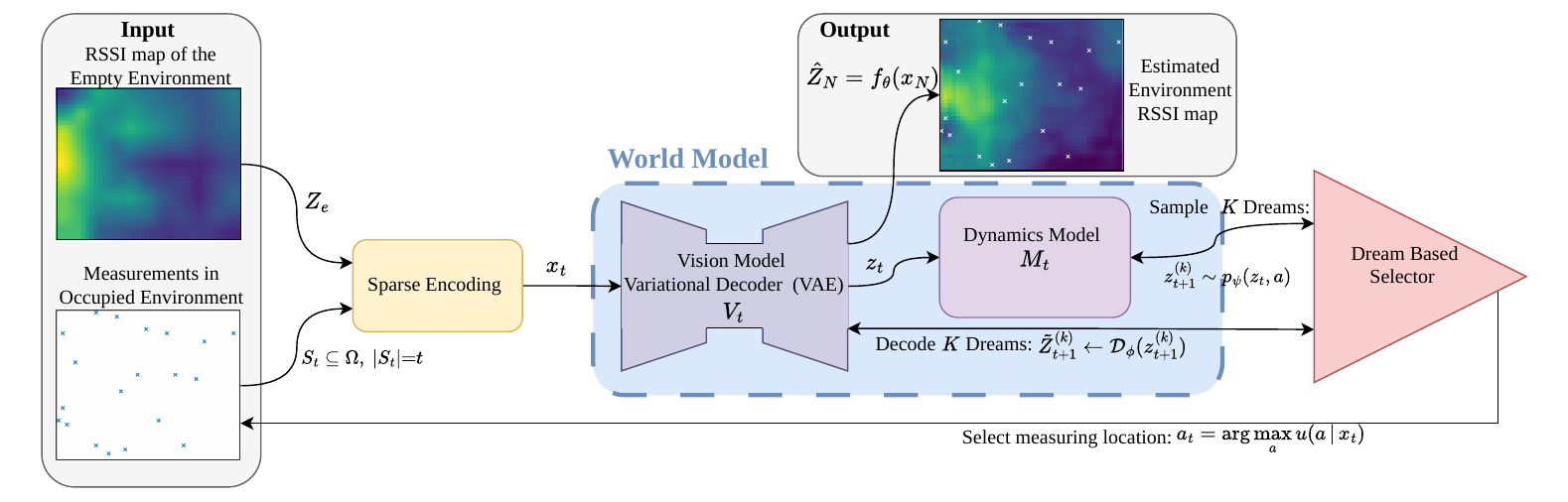}
    \caption{High-level overview of the proposed world-model–based framework for active radio environment mapping. The vision model reconstructs the occupied RSSI map from sparse measurements and an empty-environment reference map, while the dynamics model enables dreaming-based simulation of candidate measurement actions to guide measurement selection.}
    \label{fig:diagram}
    \vspace{-15pt}
\end{figure*}

World models rely on an internal representation of the environment that enables an agent to reason without directly acting in the real environment. Instead the agent relies on the concept of dreaming. Meaning, that the agent simulates the consequences of candidate actions within its learned model before executing them in the real world, allowing it to anticipate how each action may affect future observations. After an action is executed and new information about the environment is revealed, the internal representation is updated and used to refine the agent’s belief, guiding subsequent decisions in a sequential manner~\cite{hafner2025mastering}. In our work, we adapt world-model framework to active radio environment mapping.


Specifically, we propose a world-model based framework for active \ac{rssi} map reconstruction that leverages an available empty-environment map as structural prior information. As illustrated in Fig.~\ref{fig:diagram}, the framework consists of a vision model that reconstructs the occupied \ac{rssi} map from sparse observations and a dynamics model that predicts how the model’s internal belief evolves as new measurements are acquired. By simulating candidate measurement actions through dreaming, the system actively selects informative measurement locations and progressively refines its belief over the occupied environment. The final output is a dense RSSI map obtained after a limited number of measurements, i.e., obtain accurate map with only a few shots.

We validate the proposed framework using indoor WiFi RSSI data~\cite{milosheski2025indoor}. Although Wi-Fi and 6G cellular are separate technologies, both will remain foundational 
to future broadband connectivity in within the 6G ecosystem, with Wi-Fi playing a major role in enabling indoor use cases. In this context, indoor radio environment maps derived from Wi-Fi measurements capture propagation- and environment-driven characteristics (e.g., attenuation, shadowing, multipath, etc.) that are largely independent of the specific radio access technology, making the approach broadly applicable and extensible to other systems and potentially to outdoor deployments.

In summary, this paper makes the following contributions:
\begin{itemize}
\item We formulate active \ac{rssi} map reconstruction as a sequential decision-making problem and propose a world-model inspired framework to solve the problem.
\item We introduce a dreaming-based active measurement selection strategy that exploits prior information from an empty-environment \ac{rssi} map to guide sampling in occupied environments.
\item Through experimental evaluation and use of real world data\footnote{The code supporting our experiments is publicly available on GitHub: https://github.com/hribarjernej89/WorldModelsREM}, we demonstrate that the proposed framework significantly outperforms \ac{gp}–based interpolation and static baselines in the few-shot regime, achieving up to five times lower \ac{rmse} with the same number of measurements.
\end{itemize}


The rest of the paper is organized as follows. Section~\ref{sec:related} reviews related work. Section~\ref{sec:problem} presents the problem formulation, followed by the proposed solution in Section~\ref{sec:solution}. Experimental evaluation and results are presented in Section~\ref{sec:validation}, and Section~\ref{sec:conclusion} concludes the paper.





\section{Related Work}
\label{sec:related}

Our proposed solution is related to prior work that focuses on reconstructing radio maps from collected measurements using as few samples as possible~\cite{wang2020indoor, polyzos2024bayesian, lu2025bayesian, zhang2024fast, teganya2021deep}. For example, the authors in~\cite{wang2020indoor} proposed the DeepMap approach, which leverages \ac{gp} to construct accurate indoor radio environment maps from sparse WiFi \ac{rss} measurements. Similarly, the work in~\cite{polyzos2024bayesian} employs \ac{gp}-based active learning to select informative measurement locations for radio map construction, aiming to reduce the number of required samples. Bayesian neural networks combined with learning-based planning for actively selecting informative measurement locations are used in~\cite{lu2025bayesian}. The authors in~\cite{zhang2024fast} propose a \ac{gan}-based radio map estimation approach that infers spatial \ac{rss} distributions from sparse user measurements without requiring transmitter information. Lastly,~\cite{teganya2021deep} presents a deep learning–based radio map estimation method using a convolutional autoencoder to learn spatial propagation structures from prior environments and thereby reconstruct radio maps accurately.
In contrast to existing approaches that rely on interpolation, regression, or static estimates, our work adopts a world model framework in which the system learns an internal representation of the radio environment and explicitly reasons about the impact of future measurements through dream rollouts. Such a solution enables active measurement selection by predicting how the map reconstruction evolves as new samples are acquired.

\section{Problem Formulation}\label{sec:problem}

We consider the problem of active reconstruction of a \ac{rem} from sparse received \ac{rssi} measurements. The environment is discretized into a two-dimensional grid $\Omega \subset \mathbb{Z}^2$ of size $H \times W$, where each grid cell represents a candidate measurement location. For a given access point, an \emph{empty-environment} \ac{rssi} map $Z_e : \Omega \rightarrow \mathbb{R}$ is assumed to be fully observed and available. Such a reference map can be obtained during network deployment or calibration phases, when the environment is empty or in a known reference state~\cite{wilson2010radio}. The corresponding \emph{occupied} \ac{rssi} map $Z_o : \Omega \rightarrow \mathbb{R}$ is unknown and can only be queried at selected locations, i.e., grid cells. A measurement at location $i \in \Omega$ returns:

\begin{equation}
    z_i = Z_o[i].
\end{equation}

\begin{table}[t] \vspace{5pt}
\centering
\caption{List of notation used.}
\label{tab:notation}
\begin{tabular}{ll}
\hline
\textbf{Notation} & \textbf{Description} \\
\hline
$\Omega$ & Discrete grid of candidate measurement locations \\
$H,W$ & Grid height and width \\
$Z_e \in \mathbb{R}^{H\times W}$ & Empty-environment RSSI reference map \\
$Z_o \in \mathbb{R}^{H\times W}$ & Occupied RSSI map (unknown) \\
$S_t \subseteq \Omega$ & Set of measured locations at step $t$ \\
$z_i$ & RSSI measurement at location $i$, $z_i = Z_o[i]$ \\
$V_t \in \mathbb{R}^{H\times W}$ & Sparse value map of collected measurements \\
$\varepsilon(\hat{Z}_N, Z_o)$ & RMSE of reconstructed and occupied RSSI map \\
$M_t \in \{0,1\}^{H\times W}$ & Binary measurement mask \\
$x_t$  & Observation provided to the world model \\
$\pi$  & Measurement selection policy \\
$\mathcal{E}_\phi(\cdot)$ & Encoder mapping observation to latent belief \\
$z_t \in \mathbb{R}^{d_z}$ & Latent belief state at step $t$ \\
$\mathcal{D}_\phi(\cdot)$ & Decoder mapping latent state to RSSI map\\
$\hat{Z}_t$ & Reconstructed RSSI map at step $t$ \\
$a_t \in \Omega$ & Action: selected measurement location \\
$p_\psi(z_{t+1}\mid z_t,a_t)$ & Latent dynamics model \\
$\tilde{Z}_{t+1}^{(k)}$ & Dreamed RSSI map from latent rollout $k$ \\
$u(a)$ & Dreaming-based uncertainty score for action $a$ \\
$P$ & Candidate action pool size \\
$K$ & Number of dream rollouts per candidate \\
$N$ & Measurement budget \\
\hline
\end{tabular}
\end{table}

Measurement acquisition is performed sequentially. At decision step $t$, the measurements have been collected at locations $S_t \subseteq \Omega$, with $|S_t| = t$. The collected measurements are represented by a sparse value map $V_t \in \mathbb{R}^{H \times W}$ and a binary mask $M_t \in \{0,1\}^{H \times W}$, defined as:
\begin{equation}
V_t[i] =
\begin{cases}
z_i, & i \in S_t,\\
0, & i \notin S_t,
\end{cases}
\qquad
M_t[i] =
\begin{cases}
1, & i \in S_t,\\
0, & i \notin S_t.
\end{cases}
\end{equation}

\noindent Together with the empty reference map, these form the observation available to the reconstruction system:
\begin{equation}
x_t = \{ Z_e, V_t, M_t \}.
\end{equation}

Given observation $x_t$, a reconstruction model $f_\theta$ produces an estimate $\hat{Z}_t = f_\theta(x_t)$ of the occupied \ac{rssi} map. Measurement acquisition incurs cost, and the system is subject to a fixed measurement budget $N$. The objective is to select a set of measurement locations $S_N \subseteq \Omega$, with $|S_N| = N$, such that the final reconstruction
error is minimized.

Reconstruction quality is evaluated using the \ac{rmse}, denoted by $\varepsilon$, over the entire spatial domain. For a final estimate $\hat{Z}_N$ obtained after $N$ measurements, the reconstruction error is defined as:
\begin{equation}\label{eq:rmse}
\varepsilon(\hat{Z}_N, Z_o) =
\sqrt{
\frac{1}{|\Omega|}
\sum_{j \in \Omega}
\big( \hat{Z}_N[j] - Z_o[j] \big)^2
}.
\end{equation}

\noindent Minimizing $\varepsilon(\hat{Z}_N, Z_o)$ promotes accurate reconstruction of the occupied \ac{rssi} map by penalizing both large-scale deviations and localized distortions induced by environmental changes.

Under a fixed measurement budget $N$, the objective is to design a measurement selection policy $\pi$ and a reconstruction model $f_\theta$ that minimize the expected reconstruction error,
\begin{equation}\label{eq:problem}
\min_{\pi, f_\theta}
\; \mathbb{E}\!\left[
\varepsilon\!\left( \hat{Z}_N, Z_o \right)
\right]
\quad
\text{s.t.} \quad |S_N| = N,
\end{equation}
where the expectation is taken with respect to the stochasticity in measurement selection, environmental variability, and model uncertainty.

This optimization problem is inherently combinatorial, as the number of possible measurement subsets grows exponentially with grid size. Moreover, the impact of each measurement depends on previously collected measurements, making the problem sequential and history-dependent. As a result, exact optimization is computationally intractable, motivating the use of model-based strategies that reason about the expected effect of future measurements on the reconstructed \ac{rssi} map.

\section{Dreaming of Optimal Measurement Locations With World Model Framework}\label{sec:solution}

To design a solution for the problem formulated in Eq.~\ref{eq:problem}, we adopt a world model based framework, which is also illustrated in Fig.~\ref{fig:diagram}. The key idea is to reconstruct a \ac{rem} using only a small number of available \ac{rssi} measurements. As shown in Fig.~\ref{fig:diagram}, the \emph{world model} consists of two learned components: (i) a vision model, implemented as a \ac{vae}, i.e., $V_t$ , which maps the current observation to a belief environment representation and decodes it into an estimated \ac{rssi} map, and (ii) a dynamics model, i.e., $M_t$ , that predicts how this belief evolves in response to a measurement action. Active measurement selection is enabled through a dreaming mechanism that exploits these learned models. At each decision step, the system considers a set of candidate measurement locations and, for each candidate, simulates multiple imagined futures by propagating the belief through the dynamics model and decoding the resulting predicted maps.

These dreamed reconstructions represent how the system expects the reconstructed \ac{rssi} map to change if a particular measurement was taken. The variability among the dreamed reconstructions is used as a proxy for predictive uncertainty, where a lower variance indicates a more informative measurement that is expected to reduce uncertainty in the belief about the environment. Consequently, the measurement selection policy prefers actions that minimize the expected predictive variance across the dreamed outcomes. Within the proposed framework, the empty-environment \ac{rssi} map serves as a fixed structural prior that conditions the reconstruction process. Rather than being learned, this reference map is provided as part of the observation and enables the world model to exploit the underlying spatial structure of the environment before occupancy-induced perturbations are introduced. By sequentially querying \ac{rssi} values at selected grid cells in the occupied environment, the system incrementally refines its belief and predicts the full occupied \ac{rssi} map. The combination of the \ac{vae} model and the dynamics model allows the system to reason about the expected impact of future measurements and to guide measurement selection under a limited budget.

\begin{algorithm}[b!]
  \caption{Dreaming World-model solution for active measurement control for REM reconstruction}\label{alg:dream_acq}
\begin{algorithmic}[1]
\Require Empty \ac{rssi} map $Z_e$, measurement budget $N$, encoder $\mathcal{E}_\phi$, decoder $\mathcal{D}_\phi$, latent dynamics $p_\psi(z_{t+1}\mid z_t,a_t)$, candidate pool size $P$, number of dream samples $K$

\State Initialize $S_0\gets\emptyset$, $V_0\gets 0$, $M_0\gets 0$
\For{$t=0 \to N-1$}
    \State Form observation $x_t=\{Z_e,V_t,M_t\}$
    \State Encode state $z_t \gets \mathcal{E}_\phi(x_t)$
    \State Sample candidate set $\mathcal{A}_t \subseteq \Omega\setminus S_t$ with $|\mathcal{A}_t|=P$
    \For{$a \in \mathcal{A}_t$}
        \For{$k=1 \to K$}
            \State Sample dreamed map  $z_{t+1}^{(k)} \sim p_\psi(z_t,a)$
            \State Decode dreamed map $\tilde{Z}_{t+1}^{(k)} \gets \mathcal{D}_\phi(z_{t+1}^{(k)})$
        \EndFor
        \State Score $u(a)\gets \frac{1}{|\Omega|}\sum_{p\in\Omega}\mathrm{Var}_k\!\left[\tilde{Z}_{t+1}^{(k)}[p]\right]$
    \EndFor
    \State Select action $a_t\gets \arg\min_{a\in\mathcal{A}_t} u(a)$
    \State Query environment: $z_{a_t}\gets Z_o[a_t]$
    \State Update $S_{t+1}\gets S_t\cup\{a_t\}$ 
    \State Set $V_{t+1}[a_t]\gets z_{a_t}$ and $M_{t+1}[a_t]\gets 1$
\EndFor
\State Output final reconstruction $\hat{Z}_N \gets f_\theta(x_N)$ (VAE-based reconstruction from $x_N$)
\end{algorithmic}
\end{algorithm}

In Alg.~\ref{alg:dream_acq}, we outline the operation of the proposed solution. The algorithm starts by specifying the required inputs, namely the empty-environment RSSI map $Z_e$ and the measurement budget $N$, and initializing the system (line~1). At each decision step $t$, the reconstruction system receives the empty reference map $Z_e$ together with the current set of sparse occupied measurements $S_t$, encoded as a sparse value map $V_t$ and a binary mask $M_t$. These components form the observation (line~3):

\begin{equation}
    x_t = \{Z_e, V_t, M_t\}.
\end{equation}

\noindent The observation is then encoded by the \ac{vae} into a belief representation (line~4), after which a candidate set of measurement actions is sampled (line~5). For each candidate action, the algorithm performs $K$ dreaming rollouts by propagating the belief through the dynamics model and decoding the resulting predicted beliefs into RSSI maps (lines~8–9). The candidate actions are then scored using the predictive disagreement across the dreamed reconstructions, computed as the mean spatial variance (line~11):

\begin{equation}
    u(a)=\frac{1}{|\Omega|}\sum_{p\in\Omega}\mathrm{Var}_k\!\left[\tilde{Z}_{t+1}^{(k)}[p]\right].
\end{equation}

\noindent Since lower variance corresponds to a greater expected reduction in uncertainty, the next measurement location is selected as:

\begin{equation}
    a_t=\arg\min_{a\in\mathcal{A}_t}u(a)
\end{equation}

\noindent in line~13. The system then queries the occupied environment to obtain the corresponding measurement (line~14), updates the measurement set $S_{t+1}$ (line~15), and refreshes the sparse representations $V_{t+1}$ and $M_{t+1}$ (line~16). This process repeats until the measurement budget $|S_N| = N$ is exhausted. The final reconstruction is obtained by decoding the belief inferred from $x_N$, yielding the estimate $\hat{Z}_N$ (line~18).

\textbf{Implementation}: The \ac{vae} model operates on three-channel grid observations comprising the empty-environment \ac{rssi} map, sparse occupied measurements, and a binary measurement mask. Its encoder uses four $3{\times}3$ convolutional layers with ReLU activations (two with $32$ channels and two with $64$ channels), interleaved with two max-pooling layers, followed by a fully connected layer with $256$ neurons that outputs the mean and log-variance of a $64$-dimensional  state. The decoder maps this  state back to the spatial domain using a fully connected layer, two upsampling stages with convolutional and a linear layer. The dynamics model is implemented as an action-conditioned recurrent model with a single \ac{lstm} cell of $128$ hidden neurons; actions, represented as normalized spatial coordinates, are embedded via a two-layer multilayer perceptron and concatenated with the  state to predict a Gaussian distribution over the next state. During dreaming, multiple candidate actions are evaluated by sampling rollouts, decoding them into \ac{rssi} maps, and selecting the action that improve the accuracy of map reconstruction.

\section{Validation and Results}
\label{sec:validation}

\begin{figure}[b]
	\centering
	 \begin{tikzpicture}
\begin{groupplot}[group style={group name=my_plots, group size=2 by 1,horizontal sep=1.0cm}, width=0.35\textwidth],]

 \nextgroupplot[
title = (a) Training loss,
xlabel=training epoch, 
axis lines=middle,
ylabel=loss,
xmin= 0,xmax= 51,
grid=both,
minor grid style={gray!35},
xtick={10, 20, 30, 40, 50},
legend style={at={(1.7,1.35)}},
ymin=0.0,ymax=.65,
legend columns=4, 
every tick/.style={
	black,
	semithick,
}]

\addplot[style=dashed, mark size=2pt, line width=1pt, color=teal!35, smooth, mark options={solid}] table[x=epoch,y=train_loss,col sep=comma]{csv/train_results2.csv};

\addplot[style=solid, mark size=2pt, line width=1pt, color=teal!90, smooth, mark options={solid}] table[x=epoch,y=train_loss,col sep=comma]{csv/train_results.csv};

\addplot[style=dotted, mark size=2pt, line width=1pt, color=teal!65, smooth, mark options={solid}] table[x=epoch,y=train_loss,col sep=comma]{csv/train_results8.csv};





\addlegendentry{$18\times22$};
\addlegendentry{$36\times44$};
\addlegendentry{$72\times88$};

 \nextgroupplot[
	axis lines=middle,
    ylabel= \Large $\varepsilon$ \normalsize, 
    xlabel=training epoch, %
	title = (b) RMSE values, 
	grid=both,
	major grid style={gray!35},
	xtick={10, 20, 30, 40, 50},
	legend style={at={(1.29gag,0.55)}},
	xmin=0, xmax=51,
    ymin=0, ymax=3.25,
	scaled y ticks = false,
	every tick/.style={
	        black,
	        semithick,
	}]

\addplot[style=dashed, mark size=2pt, line width=1pt, color=teal!35, smooth, mark options={solid}] table[x=epoch,y=rmse, col sep=comma]{csv/train_results2.csv};

\addplot[style=solid, mark size=2pt, line width=1pt, color=teal!90, smooth, mark options={solid}] table[x=epoch,y=rmse, col sep=comma]{csv/train_results.csv};

\addplot[style=dotted, mark size=2pt, line width=1pt, color=teal!65, smooth, mark options={solid}] table[x=epoch,y=rmse, col sep=comma]{csv/train_results8.csv};




 \end{groupplot}  
 \end{tikzpicture}
	\caption{Training loss and RMSE over number of training epochs for different environment sizes.}
    \label{fig:training}
\end{figure}
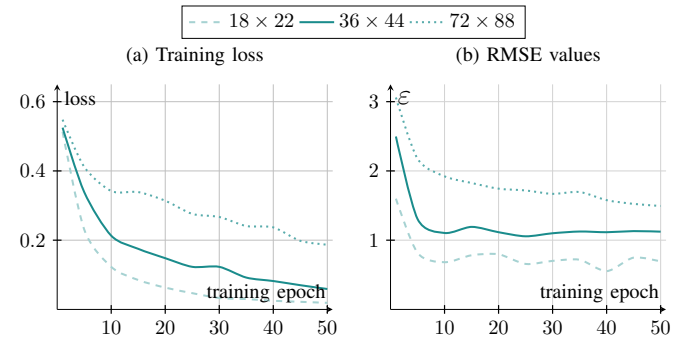

\begin{figure*}[!tbp]
  \centering
  \subfloat[Empty room map.]{\includegraphics[width=0.25\textwidth]{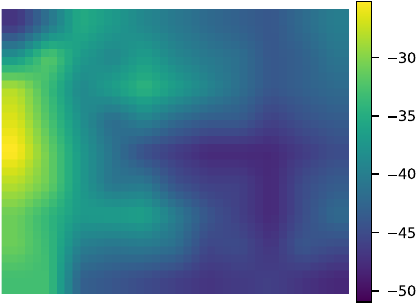}\label{fig:f1}}
  \subfloat[GP map.]{\includegraphics[width=0.25\textwidth]{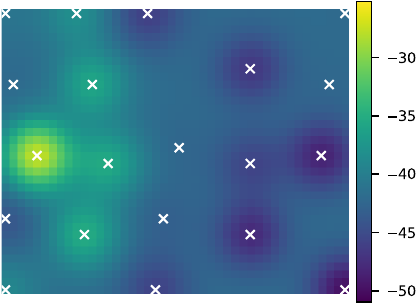}\label{fig:f2}}
  \subfloat[World model map.]{\includegraphics[width=0.25\textwidth]{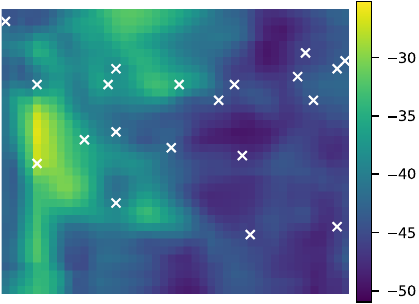}\label{fig:f2}}
  \subfloat[Ground truth map.]{\includegraphics[width=0.25\textwidth]{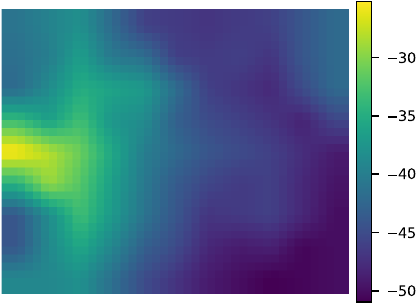}\label{fig:f2}}
  \caption{Reconstructed RSSI environment maps obtained using different methods for a grid of size $36 \times 44$ with $N=20$ measurements, with x representing selected measuring points.}
  \label{fig:heatmaps}
  \vspace{-15pt}
\end{figure*}

\begin{figure}
	\centering
	 \begin{tikzpicture}
\begin{groupplot}[group style={group name=my_plots, group size=2 by 1,horizontal sep=1.0cm}, width=0.35\textwidth],]

 \nextgroupplot[
title = (a) RMSE,
xlabel=$N$, 
axis lines=middle,
xmin= 0,xmax= 36,
ylabel=\Large $\varepsilon$ \normalsize,
grid=both,
minor grid style={gray!35},
xtick={10, 20, 30},
legend style={at={(2.1,1.45)}},
ymin=0.0,ymax=7.25,
legend columns=4, 
every tick/.style={
	black,
	semithick,
}]

\addplot[style=dashed, mark size=2pt, line width=1pt, color=gray!90, smooth, mark options={solid}] table[x=N,y=rmse_copy_empty,col sep=comma]{csv/samples_results4.csv};

\addplot[style=dotted, mark size=2pt, line width=1pt, color=brown!90, smooth, mark options={solid}] table[x=N,y=rmse_gp,col sep=comma]{csv/samples_results4.csv};

\addplot[style=solid, mark size=2pt, line width=1pt, color=teal!90, smooth, mark options={solid}] table[x=N,y=rmse_worldmodel_dream,col sep=comma]{csv/samples_results4.csv};





\addlegendentry{Empty Room Baseline};
\addlegendentry{GP};
\addlegendentry{Proposed World Model};

 \nextgroupplot[
	axis lines=middle,
    ylabel= MAE,
    xlabel= $N$, %
	title = (b) MAE, 
	grid=both,
	major grid style={gray!35},
	ytick ={1, 2, 3, 4, 5},
	xtick={10, 20, 30, 40, 50},
	legend style={at={(1.75,0.55)}},
	xmin=0, xmax=36,
    ymin=0, ymax=5.25,
	scaled y ticks = false,
	every tick/.style={
	        black,
	        semithick,
	}]

\addplot[style=dashed, mark size=2pt, line width=1pt, color=gray!90, smooth, mark options={solid}] table[x=N,y=mae_copy_empty,col sep=comma]{csv/samples_results4.csv};

\addplot[style=dotted, mark size=2pt, line width=1pt, color=brown!90, smooth, mark options={solid}] table[x=N,y=mae_gp,col sep=comma]{csv/samples_results4.csv};

\addplot[style=solid, mark size=2pt, line width=1pt, color=teal!90, smooth, mark options={solid}] table[x=N,y=mae_worldmodel_dream,col sep=comma]{csv/samples_results4.csv};




 \end{groupplot}  
 \end{tikzpicture}
	\caption{RMSE and MAE over number of samples $N$ for enviroment size $36\times44$.}
    \label{fig:samples}
	\vspace{-10pt}
\end{figure}
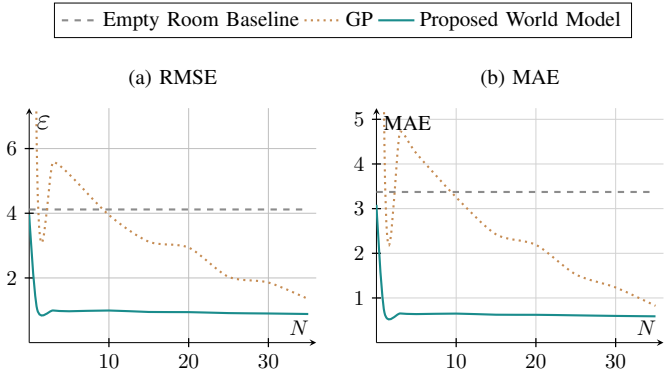

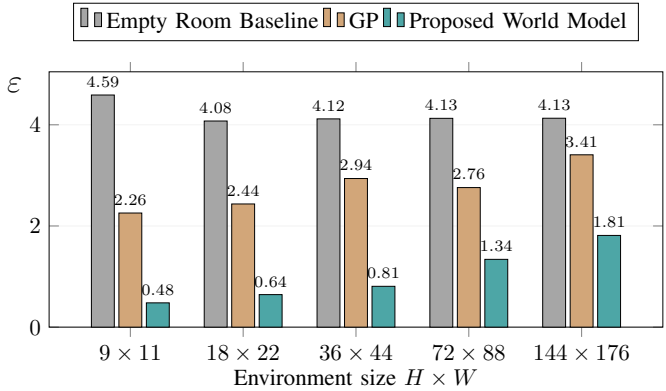
\begin{figure}
	\centering
	\begin{tikzpicture}
\begin{axis}[
    ybar,
    bar width=10pt,
    width=11cm, 
    height=5.5cm,
    ymin=0,
    ylabel={\Large $\varepsilon$ \normalsize},
    xlabel={Environment size $H \times W$},
    symbolic x coords={1,2,4,8,16},
    xtick=data,
    xticklabels={$9\times11$,$18\times22$,$36\times44$,$72\times88$,$144\times176$},
    enlarge x limits=0.18,
    legend style={at={(0.5,1.12)},anchor=south,legend columns=3},
    nodes near coords,
    nodes near coords align={vertical},
    every node near coord/.append style={font=\scriptsize},
    major grid style={opacity=0.2},
    ymajorgrids,
    group style={group size=1 by 1},
     ylabel style={
    rotate=-90,
    xshift=20pt,
    yshift=50pt},
]
\addplot[fill=gray!70] coordinates {(1,4.589) (2,4.075) (4,4.117) (8,4.128) (16,4.130)};
\addplot[fill=brown!70]  coordinates {(1,2.256) (2,2.435) (4,2.938) (8,2.759) (16,3.407)};
\addplot[fill=teal!70]  coordinates {(1,0.480) (2, 0.643) (4,0.807) (8,1.340) (16,1.813)};

\legend{Empty Room Baseline, GP, Proposed World Model}
\end{axis}
\end{tikzpicture}

	\caption{RMSE over different environment sizes for $N=20$.}
    \label{fig:barplots}
	\vspace{-15pt}
\end{figure}

In this section, we evaluate the proposed world model–based framework and compare it against two baseline methods, namely a \ac{gp}–based reconstruction and a copy empty environment map, with performance quantified using \ac{rmse} and \ac{mae}. The evaluation is conducted across multiple environment sizes, allowing us to assess robustness. 
Parameters including the candidate pool size $P$ and dream roll-outs $K$ were selected via grid search on the training scenarios. Table~\ref{tab:params} summarizes the experimental settings, including the number of training episodes $E_{\text{ep}}$ and the world-model parameters $\eta$ and $\beta_{\mathrm{KL}}$.


\textbf{Dataset}: In our validation, we employ an indoor \ac{rssi} dataset we collected from a single Wi-Fi \ac{ap} operating on 2.4 GHz band over a two-dimensional grid~\cite{milosheski2025indoor}. The dataset comprises paired measurements from an open-concept, furnished office, collected under two scenarios: unoccupied and occupied by 7-10 people performing regular working activities. Four such empty vs. occupied scenario 4 pairs were collected, where three are of them used for training the proposed framework and \ac{gp} baseline and one is reserved for validation and evaluation. As reflected in the implementation, the empty \ac{rssi} map serves as a fully observed prior, while sparse measurements from the occupied map are sequentially queried to reconstruct the full radio environment map. Furthermore, to study multi-resolution behavior, the original grid of size $9 \times 11$ grid is uniformly upscaled using bilinear interpolation with scale factors $\{1,2,4,8,16\}$, enabling evaluation over higher spatial granularities.

\begin{table}[t]
\centering
\caption{Values of experimental parameters.}
\label{tab:params}
\begin{tabular}{llll}
\hline
\textbf{Parameter} & \textbf{Value} & \textbf{Parameter} & \textbf{Value}  \\
\hline
$P$ & $40$ & $K$ & $12$ \\
$E$ & $200$ & $E_{\text{ep}}$ & $50$ \\
$\eta$ & $10^{-3}$ & $\beta_{\mathrm{KL}}$ & $10^{-3}$ \\
\hline
\end{tabular}
\vspace{-15pt}
\end{table}

\textbf{Baselines}: We employ two baseline solutions against which the performance of the proposed framework is compared. The first baseline simply copies the empty-environment \ac{rssi} map, serving as a reference to illustrate how environmental occupancy impacts the overall reconstruction performance. The second baseline employs a \ac{gp}, in which a prior model of the radio environment is constructed by learning the spatial covariance structure from the empty-environment \ac{rssi} map. Specifically, a stationary kernel composed of a constant term, a Radial Basis Function (RBF), and a white-noise component is fitted to measurements from the empty environment. The learned kernel is then reused as the prior covariance function when reconstructing the occupied-environment \ac{rssi} map. Given a set of sparse measurements from the occupied environment, the \ac{gp} posterior is obtained by conditioning this prior on the observed values, enabling interpolation over the full grid while preserving the spatial characteristics inferred from the empty environment, and thus providing a comparable solution.


We first analyze the training behavior of the proposed world model across multiple environment sizes. Fig.~\ref{fig:training} reports the training results obtained using three training scenarios for each spatial scale. Fig.~\ref{fig:training}(a) shows the training loss as a function of the number of epochs, demonstrating consistent convergence across all environment sizes. As training progresses, the loss decreases steadily, indicating that the model improves its internal representation of the radio environment and its ability to incorporate sparse measurements. On the other hand, Fig.~\ref{fig:training}(b) show the \ac{rmse} during training. While the training loss continues to decrease over epochs, the \ac{rmse} stabilizes after approximately ten epochs. This behavior suggests that, beyond this point, further optimization mainly refines the internal representation without yielding substantial improvements in reconstruction accuracy of the world map. This indicates that the world model solution reaches a stable reconstruction capability early in training.

Fig.~\ref{fig:heatmaps} presents a qualitative comparison of reconstructed \ac{rssi} environment maps obtained using different methods. The figure includes the empty-environment map, the \ac{gp}-based reconstruction, the proposed world model reconstruction, and the ground-truth occupied map. A clear discrepancy can be observed between the empty map (Fig.~\ref{fig:heatmaps}(a)) and the ground truth (Fig.~\ref{fig:heatmaps}(d)), highlighting the impact of occupancy on signal propagation in the employed dataset. Compared to the \ac{gp}-based approach (Fig.~\ref{fig:heatmaps}(b)), the proposed world model (Fig.~\ref{fig:heatmaps}(c)) produces reconstructions that are visually closer to the ground truth, particularly in regions with sparse or no measurements. 

Fig.~\ref{fig:samples} illustrates the impact of the number of available measurements $N$ on reconstruction accuracy for a fixed environment size. Both \ac{rmse} (as defined in Eq.~\ref{eq:rmse}) and \ac{mae} are reported. The empty-environment baseline remains constant across all values of $N$, serving as a reference that does not incorporate any new measurements. The proposed world model framework achieves low reconstruction error even with a very small number of measurements. Notably, with only a single measurement, the proposed framework already outperforms the GP-based approach significantly as well as the empty room baseline, as shown in Fig.~\ref{fig:samples}(a). In contrast, the GP requires a substantially larger number of samples to achieve comparable accuracy. For example, at $N=10$, the \ac{rmse} of the proposed framework is approximately four times lower than that of the \ac{gp}-based method. A similar trend is observed for \ac{mae} in Fig.~\ref{fig:samples}(b). These results indicate that the learned world model is particularly effective in the few-shot regime, while the GP gradually improves as more measurements become available.

The final evaluation, shown in Fig.~\ref{fig:barplots}, analyzes reconstruction performance across different environment sizes using bar plots. As expected, reconstruction error increases for both methods as the spatial resolution of the environment increases. However, the proposed world model consistently outperforms the GP-based approach across all scales. The relative performance gap decreases with increasing environment size. For smaller environments ($9\times11$), the proposed world model achieves up to a $4.7\times$ reduction in \ac{rmse} compared to the GP baseline. For larger environments, e.g., $144\times176$, this advantage reduces to approximately $1.9\times$.

Overall, the results highlight a clear trade-off between measurement efficiency and computational complexity. While \ac{gp}-based methods can achieve competitive performance with a sufficient number of measurements, their computational cost grows cubically with the number of observations, limiting scalability. In contrast, the proposed world model framework maintains low reconstruction error in the few-shot regime and enables efficient active measurement selection through dreaming. A key advantage of the proposed solution lies in its learned internal representation of the environment, which supports informed decision-making about where to measure next. This makes the approach particularly suitable for applications where only a small number of measurements can be collected, such as rapid radio environment mapping or user-assisted sensing scenarios.
\section{Conclusion}
\label{sec:conclusion}



This paper demonstrates the potential of world models for active radio environment mapping under limited measurement budgets. By learning an internal representation of the radio environment and using dreaming to reason about the expected impact of future measurements, the proposed framework enables informed and uncertainty-aware measurement selection without relying on dense sampling. The results show that learned world models substantially outperform GP-based interpolation in the few-shot regime, while dreaming-based acquisition further improves performance as additional measurements are collected. Beyond improved reconstruction accuracy, the key advantage of the proposed framework lies in its ability to exploit internal beliefs for decision making, which is particularly relevant for AI-native 6G systems. As future work, integrating more advanced model-based reinforcement learning techniques could further enhance long-horizon planning and adaptability in dynamic radio environments.





\section*{Acknowledgements}

This work was supported in part by the Slovenian Research Agency (ARIS) under grants P2-0016 and MN-0009. and by the Japan Society for the Promotion of Science (JSPS) under grants 23H00464, 25H01124, and 120245002.
Additional support was provided by the bilateral project MISA (BI-JP/24-26-001), funded jointly by ARIS and the JSPS.

\balance

\bibliographystyle{IEEEtran}
\bibliography{bibliography}

\end{document}